\documentclass[12pt]{article}
\usepackage{graphicx}
\usepackage{hyperref}
\usepackage{amsmath}
\usepackage[parfill]{parskip}
\usepackage{multirow}
\usepackage{tabularx}

\newcommand\pubnumber{DPF2013-208}
\newcommand\pubdate{\today}

\def\support{\footnote{Corresponding author}}

\def\Title#1{\begin{center} {\Large #1 } \end{center}}
\def\Author#1{\begin{center}{ \sc #1} \end{center}}

\newcommand\pubblock{\rightline{\begin{tabular}{l} \pubnumber\\
         \pubdate  \end{tabular}}}
\newenvironment{Abstract}{\begin{quotation}  }{\end{quotation}}
\newenvironment{Presented}{\begin{quotation} \begin{center} 
      \end{center}\bigskip 
      \begin{center}\begin{large}}{\end{large}\end{center} \end{quotation}}
\def\Acknowledgments{\bigskip  \bigskip \begin{center} \begin{large}
             \bf ACKNOWLEDGMENTS \end{large}\end{center}}

\def\beq{\begin{equation}}
\def\eeq#1{\label{#1}\end{equation}}
\def\eeqn{\end{equation}}

\def\beqa{\begin{eqnarray}}
\def\eeqa#1{\label{#1}\end{eqnarray}}
\def\eeqan{\end{eqnarray}}

\let\bar=\overbar

\def\Dslash{\not{\hbox{\kern-4pt $D$}}}
\def\dslash{\not{\hbox{\kern-2pt $\del$}}}

\def\msb{{\bar{\ssstyle M \kern -1pt S}}}

\begin{document}
\begin{titlepage}
\pubblock

\vfill
\Title{Potential Impact of a New GEM-based Detector on CMS Triggering}
\vfill

\Author{
\footnotesize
A.~Castaneda\support$^c$,
D.~Abbaneo$^a$,
M.~Abbrescia$^b$,
M.~Abi~Akl$^c$,
C.~Armaingaud$^a$,
P.~Aspell$^a$,
Y.~Assran$^d$,
S.~Bally$^a$,
Y.~Ban$^e$,
P.~Barria$^f$,
L.~Benussi$^g$,
V.~Bhopatkar$^h$,
S.~Bianco$^g$,
J.~Bos$^a$,
O.~Bouhali$^c$,
J.~Cai$^e$,
C.~Calabria$^b$,
S.~Cauwenbergh$^i$,
A.~Celik$^j$,
J.~Christiansen$^a$,
S.~Colafranceschi$^a$,
A.~Colaleo$^b$,
A.~Conde~Garcia$^a$,
G.~De~Lentdecker$^f$,
R.~De~Oliveira$^a$,
G.~de~Robertis$^b$,
S.~Dildick$^i$, 
S.~Ferry$^a$,
W.~Flanagan$^j$,
J.~Gilmore$^j$,
A.~Gutierrez$^k$,
K.~Hoepfner$^m$,
M.~Hohlmann$^h$,
T.~Kamon$^j$,
P.E.~Karchin$^k$,
V.~Khotilovich$^j$,
S.~Krutelyov$^j$,
F.~Loddo$^b$,
T.~Maerschalk$^f$,
G.~Magazzu$^l$,
M.~Maggi$^b$,
Y.~Maghrbi$^c$,
A.~Marchioro$^a$,
A.~Marinov$^a$,
J.A.~Merlin$^a$,
S.~Nuzzo$^b$,
E.~Oliveri$^a$,
B.~Philipps$^m$,
D.~Piccolo$^g$,
H.~Postema$^a$,
A.~Radi$^d$,
R.~Radogna$^b$,
G.~Raffone$^g$,
A.~Ranieri$^b$,
A.~Rodrigues$^a$,
L.~Ropelewski$^a$,
A.~Safonov$^j$,
A.~Sakharov$^k$,
S.~Salva$^i$,
G.~Saviano$^g$,
A.~Sharma$^a$,
A.~Tatarinov$^j$,
H.~Teng$^e$,
N.~Turini$^l$,
J.~Twigger$^h$,
M.~Tytgat$^i$,
M.~van~Stenis$^a$,
E.~Verhagen$^f$,
Y.~Yang$^f$,
N.~Zaganidis$^i$,
F.~Zenoni$^f$\\
\scriptsize
$^a$CERN, Geneva, Switzerland\\
$^b$Politecnico di Bari, Universit\'{a} di Bari and INFN Sezione di Bari, Bari, Italy\\
$^c$Texas A\&M University at Qatar, Doha, Qatar\\
$^d$Academy of Scientific Research and Technology, ENHEP, Cairo, Egypt\\
$^e$Peking University, Beijing, China\\
$^f$Universit\'{e} Libre de Bruxelles, Brussels, Belgium\\
$^g$Laboratori Nazionali di Frascati - INFN, Frascati, Italy\\ 
$^h$Florida Institute of Technology, Melbourne, USA\\
$^i$Ghent University, Dept. of Physics and Astronomy, Gent, Belgium\\
$^j$Texas A\&M University, College Station, USA\\
$^k$Wayne State University, Detroit, USA\\
$^l$INFN Sezione di Pisa, Pisa, Italy\\
$^m$RWTH Aachen University, III Physikalisches Institut A, Aachen, Germany\\
}
\normalsize

\vfill

\begin{Abstract}
\footnotesize
ABSTRACT: With the increases in the LHC instantaneous luminosity, maintaining effective triggering and avoiding dead time will 
become especially challenging. As the sensitivity of many physics studies, depends critically on the ability to maintain relatively low muon momentum thresholds, the identification of potential improvements in triggering is particularly important. The addition of a new muon detector to the existing CMS muon system in the very forward region allows for a substantial improvement in the performance of muon triggering. Integration of the new detector and the existing Cathode Strip Chamber system can improve the muon trigger momentum resolution due to an increase in the lever arm for the measurement of the muon bending angle. 
\end{Abstract}
\vfill
\footnotesize
\begin{Presented}
\small
DPF 2013\\
The Meeting of the American Physical Society\\
Division of Particles and Fields\\
Santa Cruz, California, August 13--17, 2013\\
\end{Presented}
\vfill
\end{titlepage}
\def\thefootnote{\fnsymbol{footnote}}
\setcounter{footnote}{0}

\section{Introduction}
The CMS muon system~\cite{cms} is designed to provide a robust, redundant and fast identification of the muons 
traversing the detector. For the initial phase of the CMS experiment, three types of gaseous detection technologies have been chosen 
according to the different background rates and magnetic field the detectors have to withstand. In the central region 
Drif Tube Chambers (DTs) are used, for the endcap region where the background rate is higher and 
the magnetic field is more intense cathode strip chambers (CSCs) are selected, in addition to assure an unambiguous 
bunch crossing identification and to build up a robust and redundant system Resistive Plate Chambers (RPCs) 
both in the barrel and endcap are used, providing a fast and accurate time measurement.
As the the LHC approaches to the the expected maximum instantaneous luminosity scenario ($\sim$10E34 cm$^{-2}$s$^{-1}$) keeping the rates under control and the momentum thresholds low enough to perform some of the 
physics searches (including some exotic Higgs signatures) represent a big challenge.  While redundancy in the muon identification is 
assured in most of the detector volume, in the very forward region $|\eta|>1.6$ the CMS muon identification relies entirely on the CSC system, which has been showing a good performance but, in the very high luminosity scenario, with about 20-40 interactions 
superimposed and with only being able to use some of the event information the trigger performance 
will degrade.  We will show that the presence of a second muon detector in that region could improve the muon reconstruction and triggering, by measuring the muon bending angle between the two detectors the rate of "mismeasured" muons can 
be reduced and at the same time a very high reconstruction efficiency can be achieved.

\section{The CMS-GEM Project}

\begin{figure}[h]
  \begin{center}
    \includegraphics[width=0.80\textwidth]{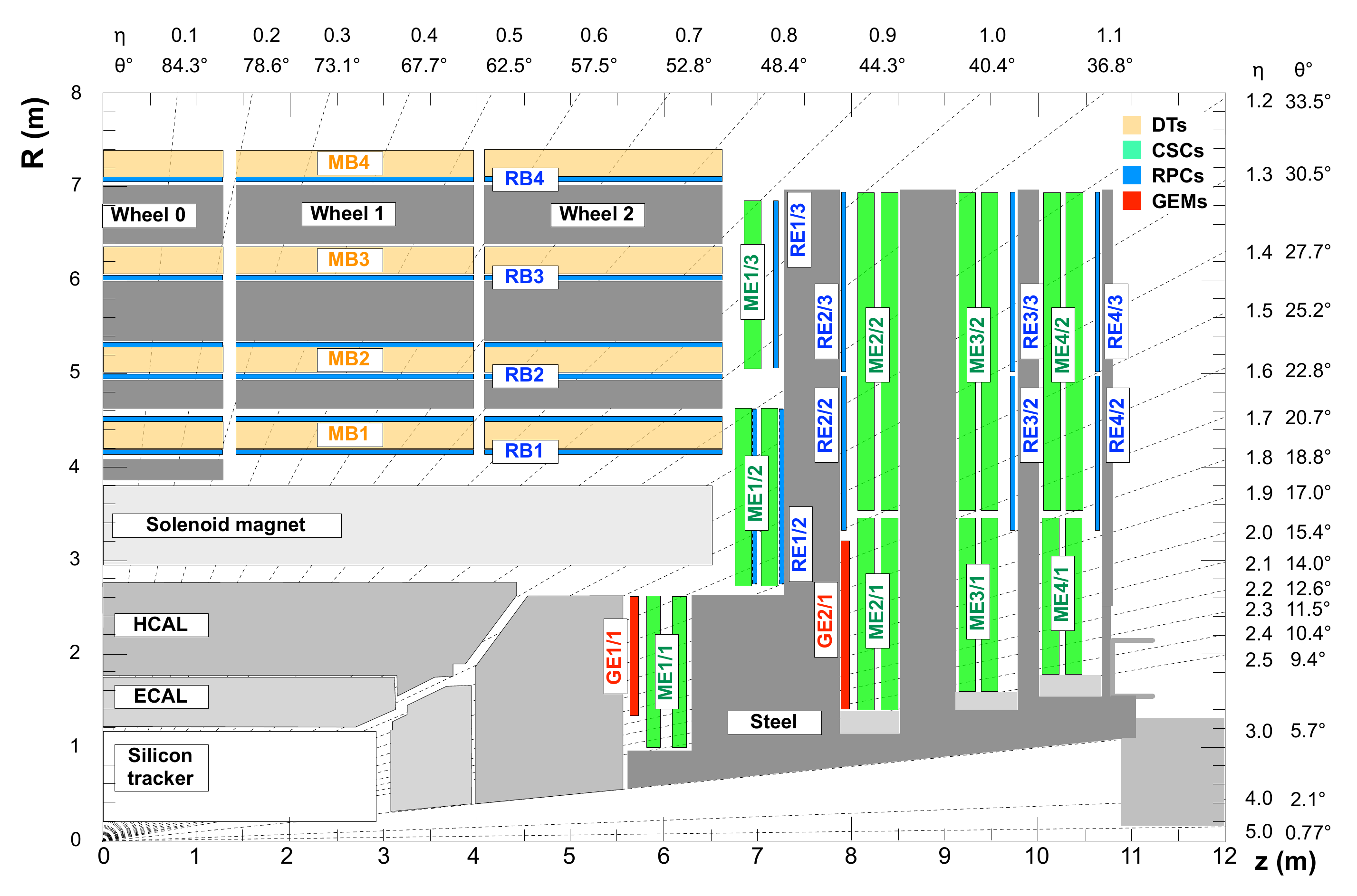}
    \caption{R-z view of the CMS detector showing the position of the different gaseous muon detectors, 
      in the barrel region: Drift Tubes (yellow) and Resistive Plate chambers (blue). 
      In the endcap region: Cathode Strip Chambers (green) and RPCs (blue) for $|\eta|<1.6$ and in the very 
      forward region (1.6$<|\eta|<$2.4) the CSCs and the proposed new GEM stations (red).}
    \label{fig:gems}
  \end{center}
\end{figure}

The CMS-GEM collaboration~\cite{gempro} has proposed the installation of two new stations of muon detector 
using Gas Electron Multiplier (GEM) technology and covering the pseudo-rapidity range between 1.6 and 2.4. 
The GEM chambers will be tentatively located just in front of the existing ME1/1 and ME2/1 CSC stations in this way they can act 
together as a combined muon identification system and extend the trigger and reconstruction capabilities. A visual 
representation of the proposed location of the GEM stations among the existing gaseous detectors is 
presented in Figure~\ref{fig:gems}. 
The GEM detector is a thin metal-coated polymer foil perforated with a high density of holes (50-100/mm$^{2}$), 
each hole acting as the multiplication region, 
single GEMs can operate up to gains of several thousands and can be used in tandem, 
the collaboration has decided to use Triple-GEM layer detector which ensure a safe operation at lower voltage. The gas volume is filled with Ar/CO$_{2}$/CF$_{4}$ mixture which was found to be the most optimal 
and the one that can offer a time resolution of 5ns with an excellent spatial resolution of the order of 100$\mu$m~\cite{gems}. 
GEM detectors are rather insensitive to ageing under sustained irradiation which implies 
that the detector could be operated without degradation at even higher integrated charges.
Single Triple-GEM chambers will be mounted face-to-face to form a double layer detector called "Super-Chamber".

One of the main purposes of the proposed GEM detector system is to provide an additional muon trigger source redundant with the 
CSC trigger to ensure robust triggering on forward muons at the high luminosity LHC scenario and beyond.  
GEM simulation assumes that the trigger is to be derived by the same or similar trigger hardware system as the current RPC system, alternatively CMS could consider designing a new L1 GEM trigger system fully tailored to GEM 
capabilities, however the RPC trigger emulation was considered as a starting point for the GEM trigger studies. 
The main goal of a muon detector is to identify high-$p_{T}$ muons produced near-by the interaction point, determine the 
bunch crossing they originate from, estimate their transverse momentum, and provide information on how good this momentum 
determination is by calculating a "muon candidate quality" figure-of-merit.  In the case of the CMS muon trigger system the 
information of every sub-detector is sent to the Global Muon Trigger (GMT) system which matches the information and decide to keep or reject the muon candidate. GEM system design and position was chosen to work together with the CSCs, the performance of the combined system depends entirely on the capabilities of 
each sub-detector, the next two sections will be devoted to review the muon identification and triggering done with 
the CSC system and how the performance will be improved by adding the GEM information.

\section{Muon identification with the CSC system}

\begin{figure}[!h]
  \begin{center}
    \includegraphics[width=0.9\textwidth]{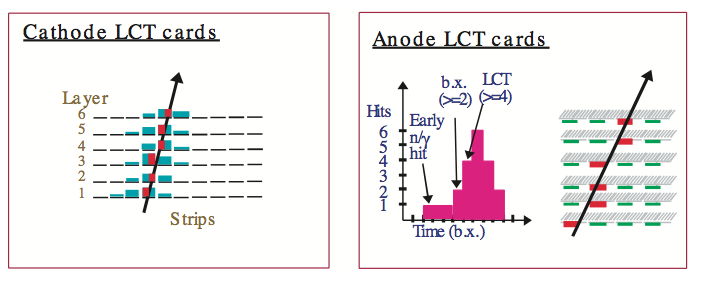}
    \caption{Principle of the CSC Local Trigger system. Anode and Cathode signals form muon segments, 
    also known as Local Charged Tracks (LCTs).}
    \label{fig:lct}
  \end{center}
\end{figure}

CSC chambers consist of arrays of positively charged "anode" wires crossed with negatively-charged cooper 
"cathode" strips within a gas volume, when muons pass through, they knock electrons off the 
gas atom, which flock the anode wires creating an avalanche of electrons. 
Because the strips and the wires are perpendicular, we get the two position coordinates for each passing
 particle.  In addition to provide precise space and time information, their closely spaced wires make the 
CSCs fast detectors suitable for triggering. Each CSC module contain six layers making it able 
to accurately identify muons and match their tracks to those in the tracker, this multilayer 
design also provide low sensitivity for the background contribution from neutrons and low energy photons.  
The way the Local CSC trigger system works is the following:  anode and cathode signals (from every 
layer in the CSC chamber) are correlated in 
time and matched to defined patterns to form muon segments, also known as Local Charged Tracks (LCTs) or "stubs"  as represented in Figure~\ref{fig:lct}. The LCTs are sent to the CSC Track Finder~\cite{trigtdr} 
which links the segments from the different endcap CSC stations, each CSC Track Finder 
can find up to three candidates that are sorted according to their quality and the best 
ones are sent to the GMT.
In a high luminosity scenario and with only the CSC system working in the very forward 
region (1.6$<\eta|<$2.4) an alternative solution to reduce the trigger rate would be to 
increase the requirement on number of "stubs" 
(up to a maximum of four), this will effectively reduce the rate but at the same time will lower the statistics 
to non-acceptable values.  The current configuration of the GMT requires to have at least three "stubs" in any of the CSC stations or two "stubs" (one of them in ME1/1).

\section{GE1/1-ME1/1 combined system and the measurement of the muon bending angle}
\begin{figure}[h]
  \begin{center}
    \includegraphics[width=0.48\textwidth]{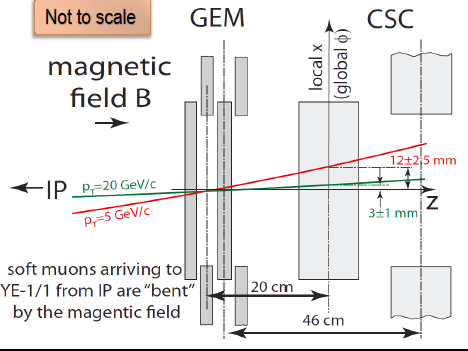}
    \includegraphics[width=0.48\textwidth]{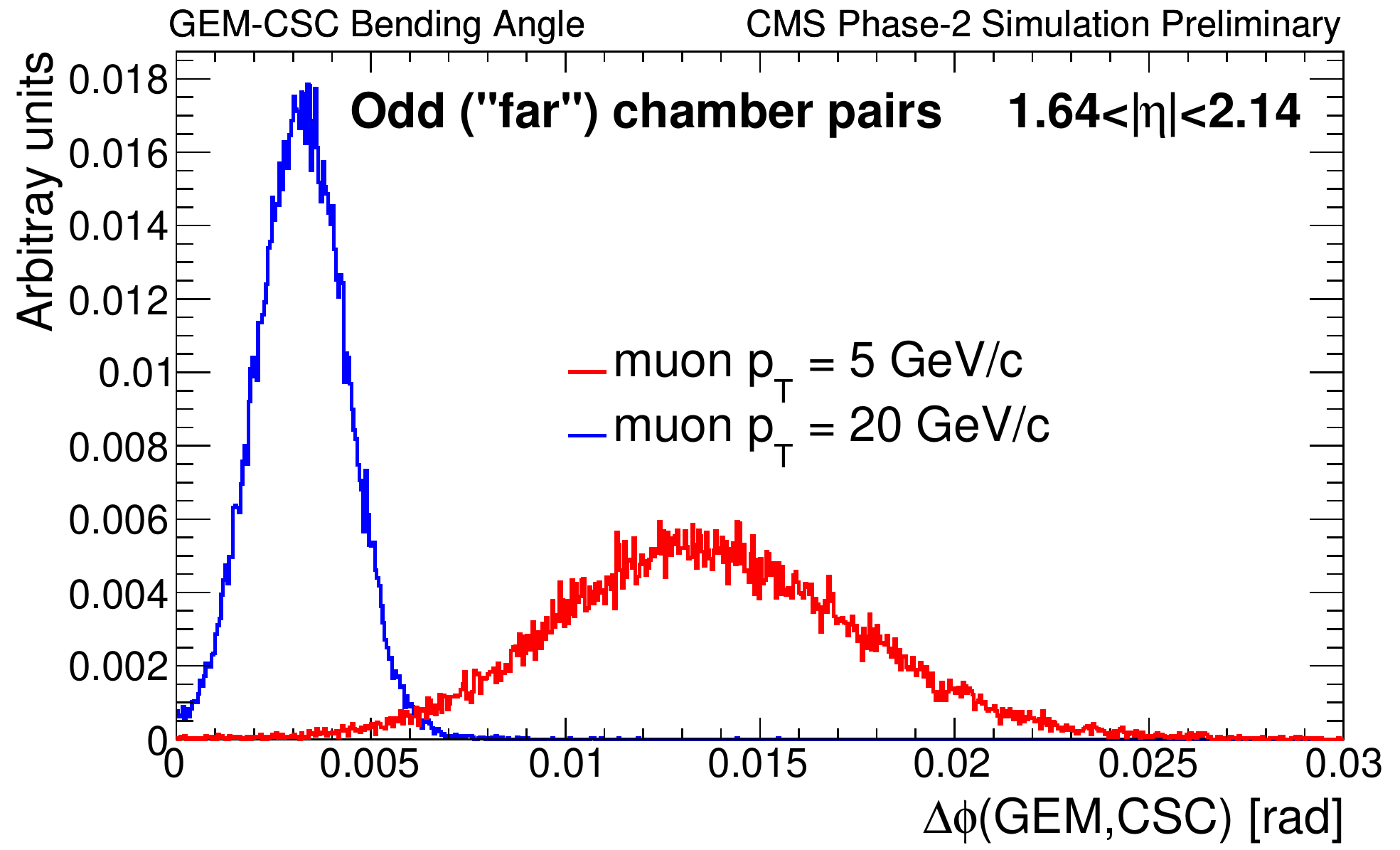}
    \caption{Left: Visualization of the GEM and CSC chambers and the muon bending angle for a low and high-$p_{T}$ muon track, 
      right: The bending angle distribution measured between GE1/1 and ME1/1 stations, the plot was produced using two 
      Monte Carlo samples for low and high-$p_{T}$ muons, the discrimination power between a 5~GeV and a 20~GeV muon is evident.}
    \label{fig:bending}
  \end{center}
\end{figure}

\begin{figure}[h]
  \begin{center}
    \includegraphics[width=0.68\textwidth]{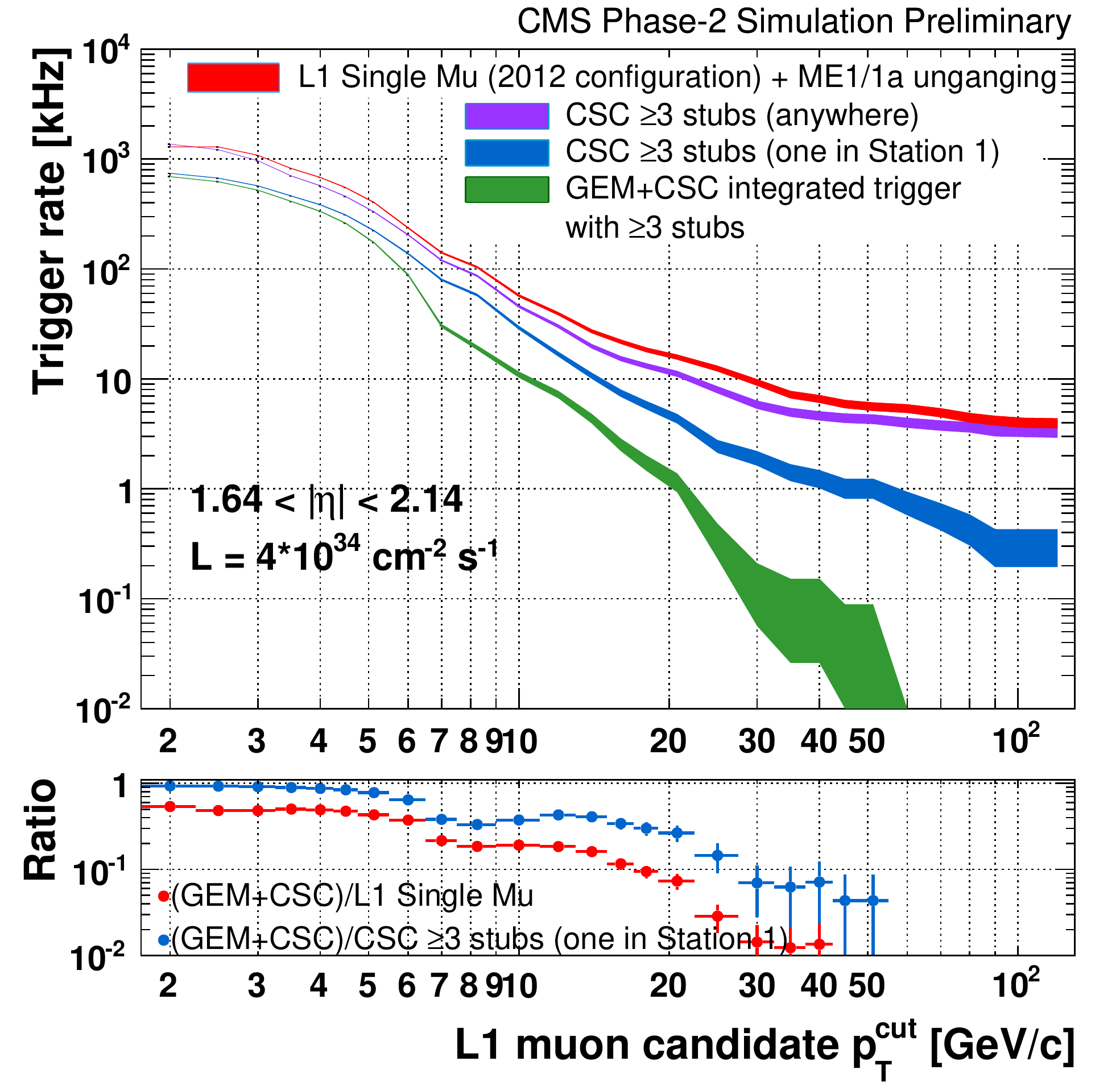}
    \caption{Comparison of the trigger rates for the Global Muon Trigger in the 2012 configuration with at least 3 stubs (loose), at least 3 stubs with at least one 1 stub from ME1/b (medium), and at least one 3 stubs with at least one 1 stub from ME1/b and a GEM pad signal (tight). The bottom plot shows the ratio of tight/GMT and tight/medium}
    \label{fig:rate}
  \end{center}
\end{figure}

The innermost CSC station (ME1/1) has a special relevance due to the fact that the 
magnetic field is more intense in this region, therefore the muon resolution is highly driven by this station. This was one of the 
motivations for the GEM collaboration to start the project with the installation of a GEM station (GE1/1) in front of the of 
the existing CSC ME1/1 station. One of the advantage of having two muon detectors within a small distance and not affected by any shielding between them is the possibility to measure with good accuracy the bending angle between the two detectors as shown in Figure~\ref{fig:bending}.
GEM signals are matched with the muon candidates from the CSC system for a specific bunch crossing and from the position in the two chambers we are able to calculate the bending. 
This variable has been measured using a full CMS Monte Carlo Simulation, a set of Monte Carlo Minimum Bias 
and Muon Gun samples were generated to study rates and efficiency respectively, from the results it
is clear that the bending angle distribution provides a powerful tool to discriminate between 
low and high-$p_{T}$ muons as shown in Figure~\ref{fig:bending}, this extra information can be used by 
the trigger system to reduce the rate of mismeasured  muons and with this also the trigger 
rate offering the the possibility of lowering the muon momentum thresholds.
In the Figure~\ref{fig:rate} it is shown the improvement in the rate by using the combined GEM-CSC trigger system compared with different scenarios of the only CSC system, 
the results presented were obtained assuming a luminosity scenario of $4\times10^{34}$cm$^{-2}$s$^{-1}$. 
Even in the tighter CSC only scenario the addition of GEM information further reduce the rate, for instance, 
for muons with $p_{T}>25$~GeV the rate is reduced by a factor of five.

\section{Summary and Outlook}

The CMS-GEM collaboration has proposed the installation of two stations of Gaseous Electron Multiplier (GEM) detectors 
in the very forward region of the CMS detector ($|\eta|>1.6$), this new detector technology is designed to work together 
with the existing CSC chambers to provide a redundant and robust identification of muons.  From Monte Carlo simulation 
studies it has been shown that in the high LHC luminosity scenario having two muon detectors near-by will help to reduce 
the trigger rate and as a consequence the possibility of lowering then muon the $p_{T}$ thresholds, if we are 
able to measure with good accuracy the muon bending angle between the two detectors.

\Acknowledgments
The corresponding author is supported by the Qatar National Research Fund (QNRF) under project NPRP-5-464-1-080

\end{document}